**Taj Dini Mahyar, V. Sokolov**

# PENETRATION TESTS FOR BLUETOOTH LOW ENERGY AND ZIGBEE USING THE SOFTWARE DEFINED RADIO

This article discusses the available Software Defined Radios (SDRs), compatible software, message formats, and also shows how it is possible to do penetration tests using SDR for Bluetooth Low Energy (BLE) and ZigBee technologies.

With the exponential growth in the ways and means by which people need to communicate: data communications, voice communications, video communications, broadcast messaging, command and control communications, emergency response communications, etc. modifying radio devices easily and cost-effectively has become business critical. Software defined radio (SDR) technology brings the flexibility, cost efficiency and power to drive communications forward, with wide-reaching benefits realized by service providers and product developers through to end users.

A number of definitions can be found to describe Software Defined Radio, also known as Software Radio or SDR. The Wireless Innovation Forum, working in collaboration with IEEE P1900.1 group, has worked to establish a definition of SDR that provides consistency and a clear overview of the technology and its associated benefits [1].

A radio is any kind of device that wirelessly transmits or receives signals in the radio frequency (RF) part of the electromagnetic spectrum to facilitate the transfer of information. In today's world, radios exist in a multitude of items such as cell phones, computers, car door openers, vehicles, and televisions.

Traditional hardware based radio devices limit cross-functionality and can only be modified through physical intervention [2; 3]. This results in higher production costs and minimal flexibility in supporting multiple waveform standards. By contrast, software defined radio technology provides an efficient and comparatively inexpensive solution to this problem, allowing multi-mode, multi-band and/or multi-functional wireless devices that can be enhanced using software upgrades (see Table 1).

SDR defines a collection of hardware and software technologies where some or all of the radio's operating functions (also referred to as physical layer processing) are implemented through modifiable software or firmware operating on programmable processing technologies. These devices include Field Programmable Gate Arrays (FPGA), Digital Signal Processors (DSP), General Purpose Processors (GPP), programmable System on Chip (SoC) or other application specific programmable processors. The use of these technologies allows new wireless features and capabilities to be added to existing radio systems without requiring new hardware.

For each SDR we compare the cost, frequency range, ADC resolution, maximum instantaneous bandwidth (see table 1), whether or not it can transmit and if it has any pre-selectors built in. Here is a quick list to what some of these metrics mean:

− frequency range — the range of frequencies the SDR can tune to;
− pre-selectors — analogue filters on the front end to help reduce out of band interference and imaging;
− instantaneous bandwidth — the size of the real time RF chunk available;
− communication — can the radio receive and/or transmit;
− ADC resolution — more resolution means more dynamic range, less signal imaging, a lower noise floor, more sensitivity when strong signals are present and better ability to discern weak signals. Some SDR's give their resolution in Effective Number of Bits (ENOB) which stands for effective number of bits.

*Table 1. Comparison of SDRs*

| Name | Cost, $ | Frequency range, MHz | ADC resolution, bits | Max bandwidth, MHz | Communication | Pre-selectors | Release date |
|---|---|---|---|---|---|---|---|
| R820T RTL2832U | 10–22 | 24–1766 | 8 | 3.2/2.4 | RX only | R820T | Aug'16 |
| Airspy R2 | 169 | 24–1750 | 12 | 10 | RX only | R820T | Nov'14 |
| HackRF One | 299 | 1–6000 | 8 | 20 | half-duplex | none | Apr'14 |
| LimeSDR | 299 | 0.1–3800 | 12 | 61.44 | full-duplex | none | Apr'16 |
| BladeRF | 420–650 | 300–3800 | 12 | 28 | full-duplex | none | Jul'13 |

**HackRF One Advantages**

Penetration testers working with wireless signal security face a wide selection of tools for testing common radio frequencies. Because all radio devices vary in effectiveness at handling frequency ranges and signal types, penetration testers must ensure the device meets their RF spectrum needs. We determined the HackRF One to be the best bang for your buck.

The HackRF One has robust features, but most of all it is well supported by a variety of open-source software on most standard computer platforms. Many newcomers enter RF penetration testing using the R820T RTL2832U or RTL-SDR. The RTL-SDR is a very inexpensive device but has several limitations. While the HackRF One costs almost $300 it also offers more capability than lower cost devices. Most notable are the features are wide frequency range and up to 20 million samples per second.

Then simply because price and wide range of supported frequency just with one disadvantage that it's half-duplex but in Our Experiment it doesn't matter then we are choosing HackRF One as our SDR hardware (Fig. 1).

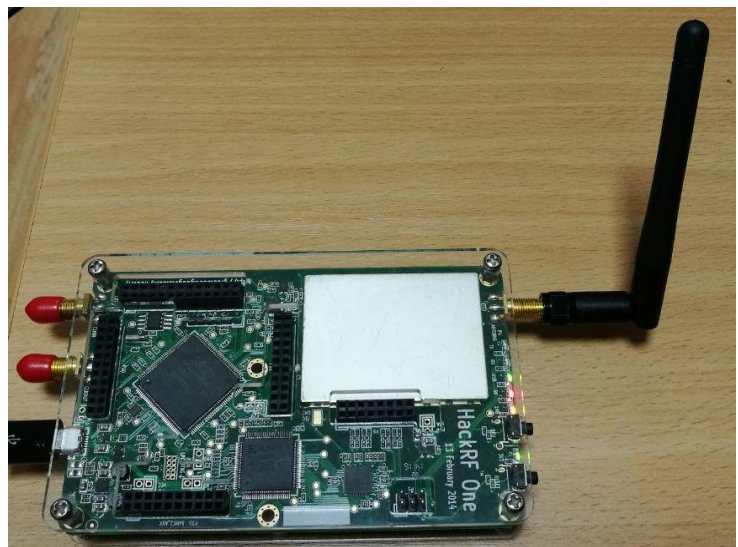

Fig. 1. General view of the used HackRF One

**SDR's Tools Discovery**

There are now dozens of software defined radio packages that support the very cheap RTL-SDR. On this paper we will attempt to list, categorize and provide a brief overview of each software program. We categorize the programs into general purpose software, single purpose software, research software and software compatible with audio piping. But here we used just a few of them as we list below (Table 1).

SDR# (pronounced "SDR sharp") is the most popular free RTL-SDR compatible software in use at the moment for OS Windows. It is relatively simple to use compared to other SDR software and has a simple set up procedure. SDR# is a simple to use program that also has some advanced features. It has a useful modular plugin type architecture, and many plugins have already been developed by third party developers. The basic SDR# download without any third party plugins includes a standard FFT display and waterfall, a frequency manager, recording plugin and a digital noise reduction plugin. SDR# also decodes RDS signals from broadcast FM [4; 5].

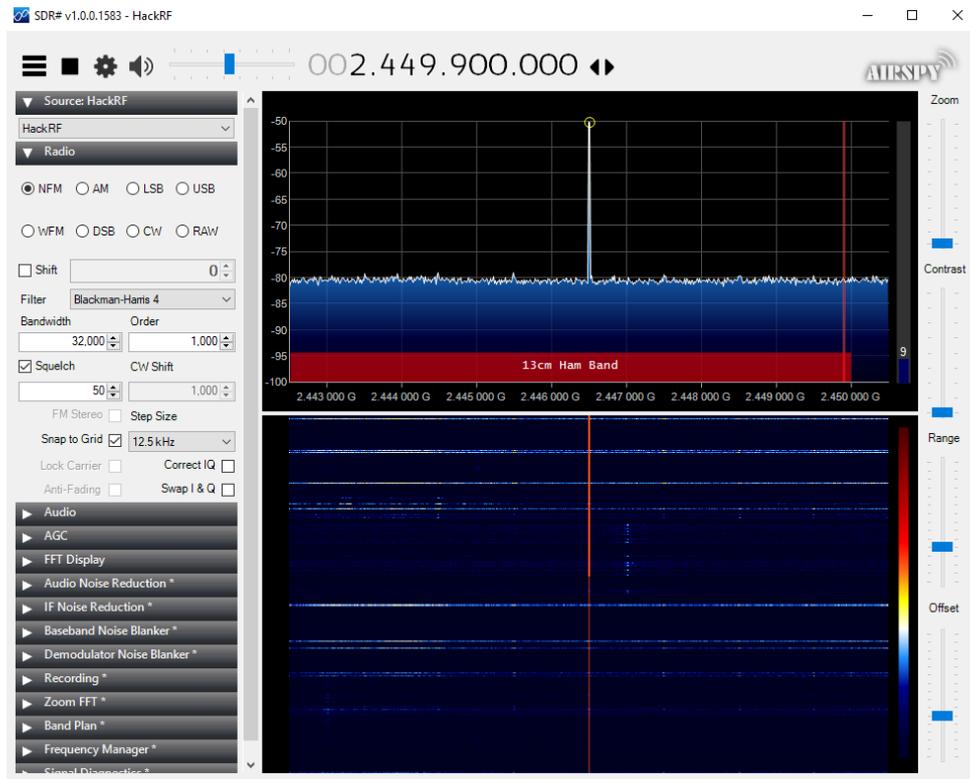

Fig. 2. SDR# main window

GQRX is a free simple to use SDR receiver, which runs on Linux and Mac systems. It is similar to SDR# in terms of features and simplicity of use. GQRX comes with a standard FFT spectrum and waterfall display and a number of common filter settings (Fig. 3).

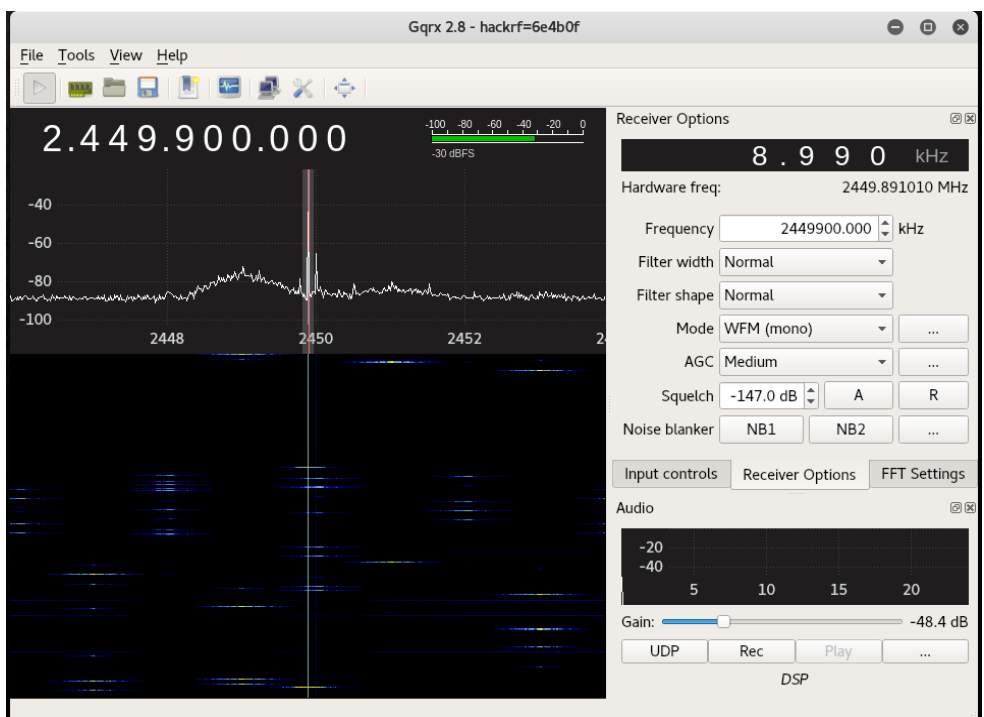

Fig. 3. GQRX main window

GNU Radio is a free and open-source software development toolkit that provides signal processing blocks to implement software radios. It can be used with readily-available low-cost external RF hardware to create software-defined radios, or without hardware in a simulation-like environment. It is widely used in research, industry, academia, government, and hobbyist environments to support both wireless communications research and real-world radio systems.

GNU Radio performs all the signal processing. You can use it to write applications to receive and transmit data with radio hardware, or to create entirely simulation-based applications. GNU Radio has filters, channel codes, synchronization elements, equalizers, demodulators, vocoders, decoders, and many other types of blocks, which are typically found in signal processing systems. More importantly, it includes a method of connecting these blocks and then manages how data is passed from one block to another. Extending GNU Radio is also quite easy; if you find a specific block that is missing, you can quickly create and add it.

GNU Radio applications can be written in either C++ or Python programming language, while the performance-critical signal processing path is implemented in C++ using processor floating-point extensions where available. This enables the developer to implement real-time, high-throughput radio systems in a simple-to-use, rapid-application-development environment.

**BLE Penetration Tests**

To capture BLE Frequency Hopping Link we need pass few steps shown on Fig. 4.

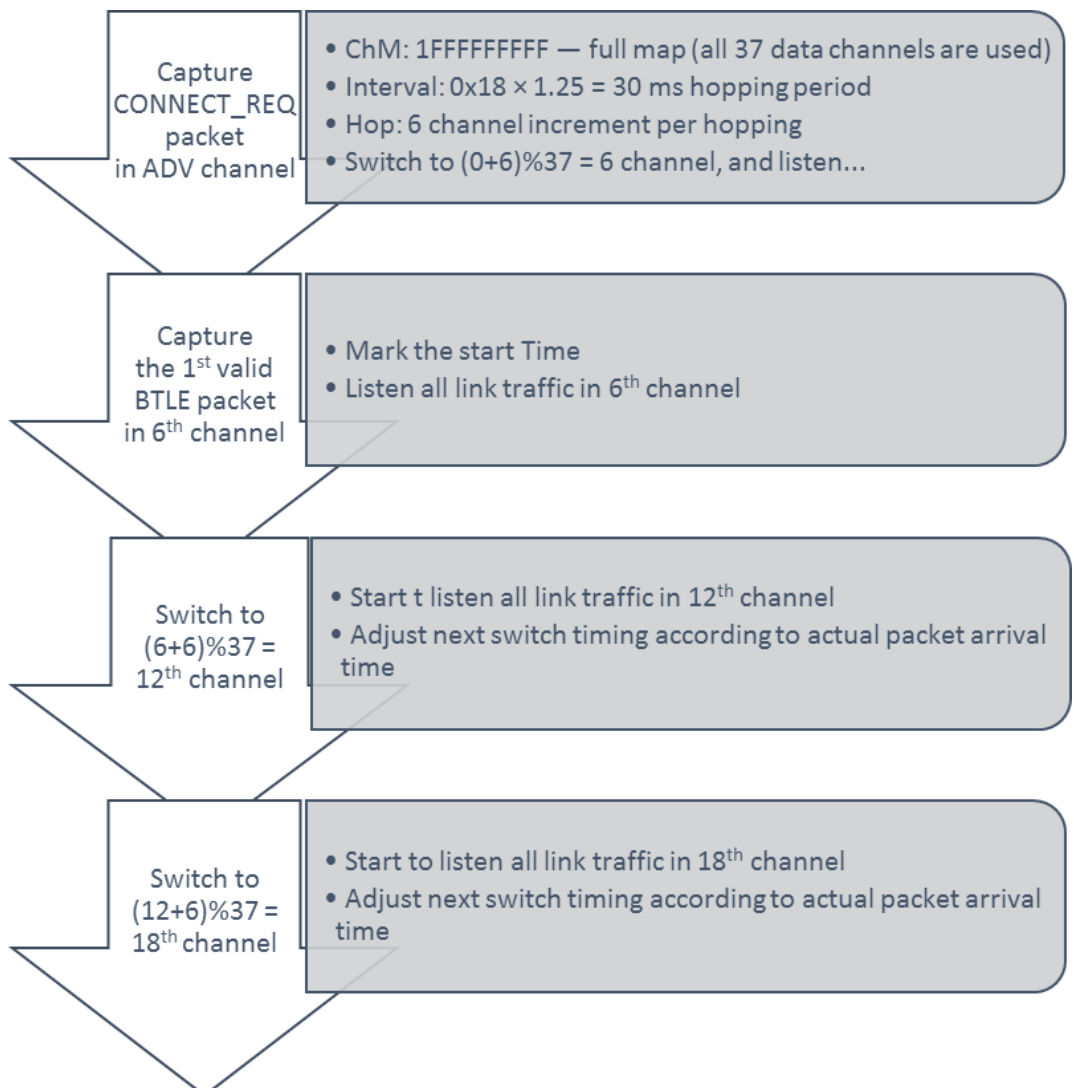

Fig. 4. Steps for frequency hopping link for BLE

We created a iBeacon packet from phone with the URL as a data inside, the packet, on channel 37 and broadcast it, after running `./BLE_rx -g 0` the result shown as below:

```
163342us Pkt192 Ch37 AA:8e89bed6 ADV_PDU_t2:ADV_NONCONN_IND T1 R0
PloadL25 AdvA:41e0302e6669
Data:0303aafe0e16aafe10bb0074616a64696e690a CRC0
```

iBeacons are certainly a trending topic recently. They allow indoor positioning, letting your phone know that you are in range of a beacon. This can have many applications: from helping you to find your car in a parking garage, through coupons and location-aware special offers in retail, to a whole lot of apps that we can't imagine right now.

Bluetooth Low Energy is a part of the Bluetooth 4.0 specification, which was released back in 2010. It originated in 2006 in Nokia as Wibree, but has since been merged into Bluetooth. It is a different set of protocols than "classic" Bluetooth, and devices are not backwards-compatible. Hence, you can now encounter three type of devices:
– Bluetooth: supporting only the "classic" mode;
– Bluetooth Smart Ready: supporting both "classic" and LE modes;
– Bluetooth Smart: supporting only the LE mode.

Newer smartphones (iPhone 4S+, SG3+), laptops, tablets, are all equipped with full Bluetooth 4.0 and hence "Smart Ready". Beacons, on the other hand, only support the low energy protocols (which allows them to work on a single battery for a long time) and hence they implement "Bluetooth Smart". Older devices, like peripherals, car systems, and older phones usually support only the classic Bluetooth protocol.

The focus in BLE is of course low energy consumption. For example, some beacons can transmit a signal for 2 years on a single cell battery (the batteries are usually not replaceable; you will probably just replace the beacon when they stop working). Both "classic" and BLE use the same spectrum range (2.4000–2.4835 GHz). The BLE protocol has lower transfer rates, however it is not meant to stream a lot of data, but rather for discovery and simple communication. In terms on range, both LE and "classic" Bluetooth signal can reach up to 100 meters.

BLE communication consists of two main parts: advertising and connecting.

Advertising is a one-way discovery mechanism. Devices which want to be discovered can transmit packets of data in intervals from 20 to 10 000 milliseconds. The shorter the interval, the shorter the battery life, but the faster the device can be discovered. The packets can be up to 47 bytes in length and consist of:
– preamble — 1 byte;
– access address — 4 bytes;
– advertising channel PDU — 2–39 bytes;
– CRC — 3 bytes.

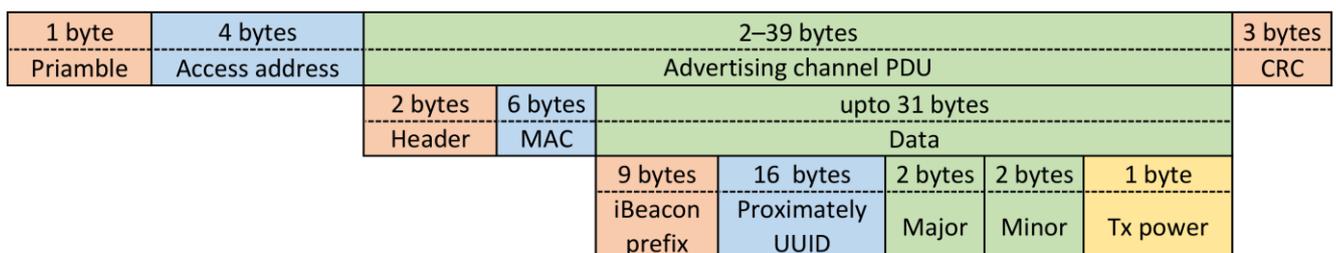

Fig. 5. BLE packet format

For advertisement communication channels, the access address is always 0x8E89BED6. For data channels, it is different for each connection.

The PDU in turn has its own header (2 bytes: size of the payload and its type — whether the device supports connections, etc.) and the actual payload (up to 37 bytes). Finally, the first 6 bytes of the payload are the MAC address of the device, and the actual information can have up to 31 bytes.

BLE devices can operate in a non-connectable advertisement-only mode (where all the information is contained in the advertisement), but they can also allow connections (and usually do).

After a device is discovered, a connection can be established. It is then possible to read the services that a BLE device offers, and for each service its characteristics (this is also known as an implementation of a GATT profile). Each characteristic provides some value, which can be read, written, or both. For example, a smart thermostat can expose one service for getting the current temperature/humidity readings (as characteristics of that service) and another service and characteristic to set the desired temperature. However, as beacons don't use connections [6].

**Breaking Down the iBeacon Format**

Apart from the mostly fixed iBeacon prefix (02 01 … 15). The proximity UUID (B9 … 6D in our example), is an identifier which should be used to distinguish your beacons from others. If, for example, beacons where used to present special offers to customers in a chain of stores, all beacons belonging to the chain would have the same proximity UUID. The dedicated iPhone application for that chain would scan in the background for beacons with the given UUID.

The major number (2 bytes, here: 0x0049, so 73) is used to group a related set of beacons. For example, all beacons in a store will have the same major number. That way the application will know in which specific store the customer is.

The minor number (again 2 bytes, here: 0x000A, so 10) is used to identify individual beacons. Each beacon in a store will have a different minor number, so that you know where the customer is exactly.

**ZigBee Penetration Tests**

In our experiment, we used Pololu Wixel module as ZigBee Rx/Tx to send one packet per second with static data as we programmed in our case FFEEFFEE on 2.4499 GHz.

The Pololu Wixel is a general-purpose programmable module featuring a 2.4 GHz radio and USB. The Wixel is based on the CC2511-F32 microcontroller from Texas Instruments, which has an integrated radio transceiver, 32 kB of flash memory, 4 kB of RAM, and a full-speed USB interface. A total of 15 general-purpose I/O lines are available, including 6 analog inputs, this small Module working on 2400–2483.5 MHz frequency range. It has approximately 50 feet under typical condition indoor range and up-to 350 kb/s programmable bit rate, and 10 kb/s effective data rate.

As mentioned in the ZigBee protocol stack that ZigBee MAC layer frame composed of MAC header, MAC payload and FCS. The diagram below depicts generic mac frame format adopted in ZigBee technology at MAC layer. This part is also referred as MPDU or MAC protocol data unit. This is embedded into PPDU (physical PDU) frame of ZigBee [12].

Without interruption, we can send and receive packets with just about 7% errors as it shows in Fig. 6. Then we tried to cache these signals with HackRF One and resend them, like cloning the packets. For this, we designed a simple diagram as in diagram shown (Fig. 7) in GNU Radio to cache signals and save them in binary file.

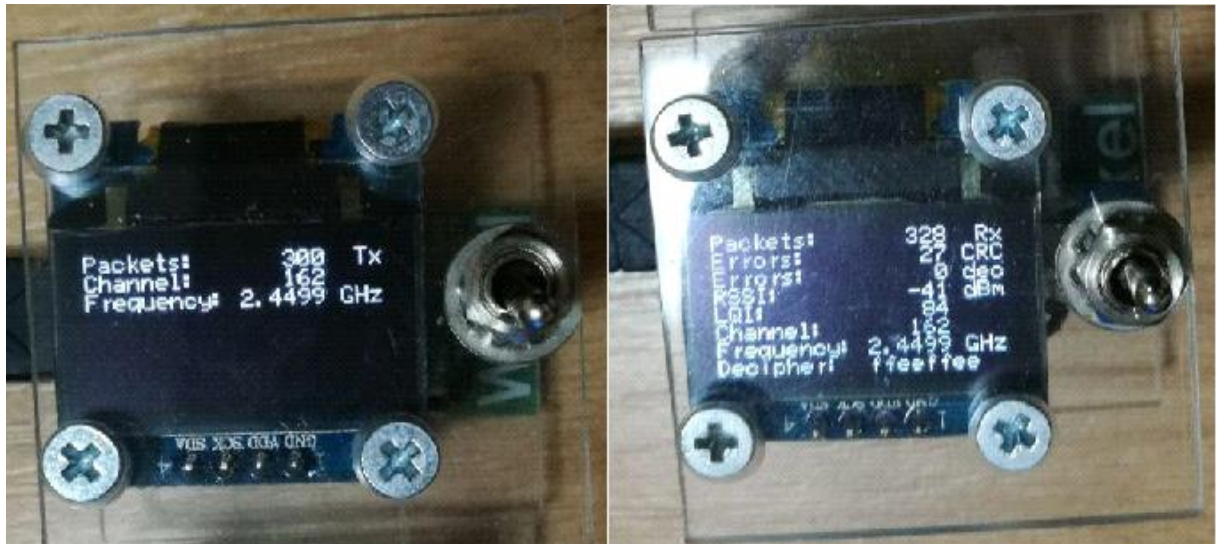

Fig. 6. Hardware receiver and transmitter on Pololu Wixel model with OLED

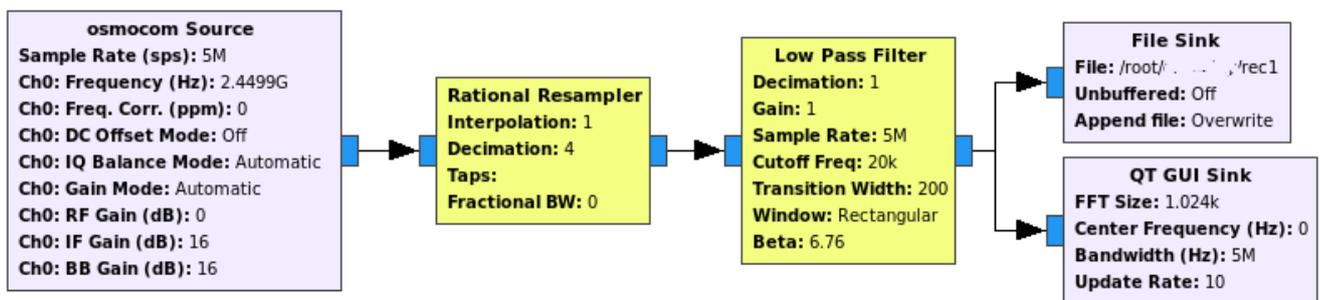

Fig. 7. GNU Radio scheme of receiver

As it shows the result after file sink will be "File" path with the name of "rec1" then we created a transceiver to send exactly same signals as we cached in Binary File. The scheme of this transmitter shown below (Fig. 8).

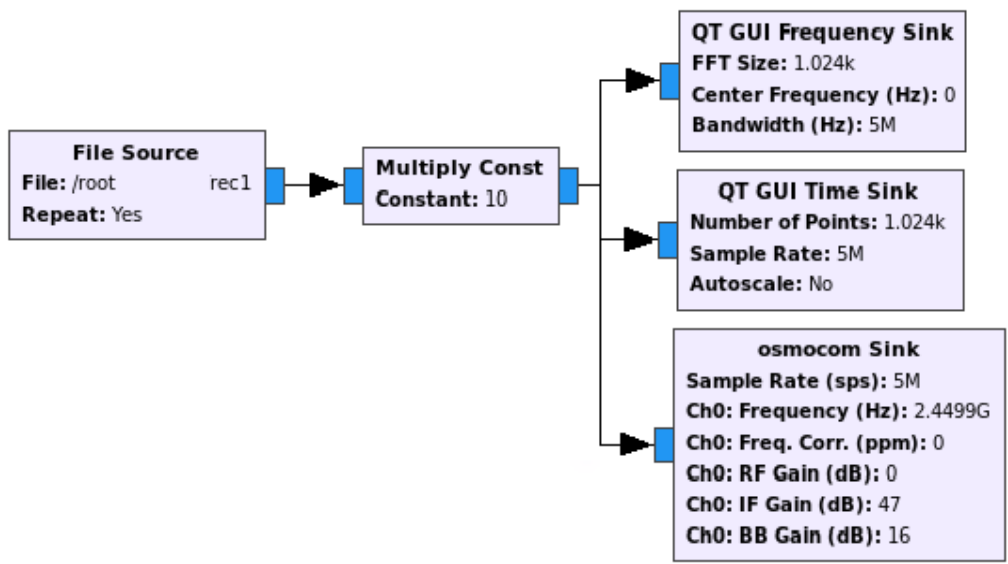

Fig. 8. GNU Radio scheme of transmitter

In this experiment, we received more than 50% of signals without error it means in real communication we had about 7% packet lost and after sniff them with HackRF One, and resend them, we had 43% more packet lost, by the way its good statistics for such experiment because it's possible to jamming signals a few times and cache. For example, authentication packets and resend them for authenticator or in other experiment we cached car remote control signals it was on 433 MHz and by same ratio we could open and close the door without real remote control key.

The use of SDR is advisable for conducting experiments with attacks on penetration, as well as for emulating receivers and transmitters for Bluetooth Low Energy and ZigBee protocols.